\documentstyle[twoside,psfig]{article}
\input{ijmpb-macros}

\begin{document}

\runninghead{A survey of nonadiabatic superconductivity
in cuprates and fullerides} {A survey of nonadiabatic superconductivity
in cuprates and fullerides}

\normalsize\textlineskip
\thispagestyle{empty}
\setcounter{page}{1}

\copyrightheading{}			

\vspace*{0.88truein}

\fpage{1}
\centerline{\bf A SURVEY OF NONADIABATIC SUPERCONDUCTIVITY}
\vspace*{0.035truein}
\centerline{\bf IN CUPRATES AND FULLERIDES}
\vspace*{0.37truein}

\centerline{\footnotesize EMMANUELE CAPPELLUTI}
\vspace*{0.015truein}
\centerline{\footnotesize\it Dipartimento di Fisica, 
Universit\`a ``La Sapienza'', P.le Aldo Moro 2}
\baselineskip=10pt
\centerline{\footnotesize\it Roma, 00185, Italy} 
\baselineskip=10pt
\centerline{\footnotesize\it and INFM, Unit\`a Roma1}

\vspace*{10pt}
\centerline{\footnotesize CLAUDIO GRIMALDI}
\vspace*{0.015truein}
\centerline{\footnotesize\it \'Ecole Polytechnique F\'ed\'erale, 
D\'epartment de microtechnique IPM}
\baselineskip=10pt
\centerline{\footnotesize\it Lausanne, CH-1015, Switzerland}

\vspace*{10pt}
\centerline{\footnotesize LUCIANO PIETRONERO}
\vspace*{0.015truein}
\centerline{\footnotesize\it Dipartimento di Fisica, 
Universit\`a ``La Sapienza'', P.le Aldo Moro 2}
\baselineskip=10pt
\centerline{\footnotesize\it Roma, 00185, Italy} 
\baselineskip=10pt
\centerline{\footnotesize\it and INFM, Unit\`a Roma1}

\vspace*{10pt}
\centerline{\normalsize and}

\vspace*{10pt}
\centerline{\footnotesize SIGFRID STR\"ASSLER}
\vspace*{0.015truein}
\centerline{\footnotesize\it \'Ecole Polytechnique F\'ed\'erale, 
D\'epartment de microtechnique IPM}
\baselineskip=10pt
\centerline{\footnotesize\it Lausanne, CH-1015, Switzerland}
\vspace*{0.225truein}

\abstracts{High-$T_c$ superconductors are characterized
by very low carrier densities.
This feature leads to two fundamental consequences:
on one hand the Fermi energies are correspondingly small and they can be of
the same order of phonon frequencies. In such a situation nonadiabatic 
corrections arising from the breakdown of Migdal's theorem
can not be longer neglected.
In addition, small carrier densities imply poor screening and correlation
effects have to be taken into account.
We present a comprehensive overview of the theory of superconductivity
generalized into the nonadiabatic regime which is qualitatively
different from the conventional one.
In this framework some of the  observed properties of the cuprates
and the fullerene compounds
can be naturally accounted for, and a number of theoretical
predictions are proposed that can be experimentally tested.}{}{}

\textlineskip			
\vspace*{12pt}			

\textheight=7.8truein
\setcounter{footnote}{0}
\renewcommand{\thefootnote}{\alph{footnote}}

After almost $15$ years and 
in spite of the huge amount of work spent in the field,
no definitive description of the high temperature superconductivity 
phenomenon has been still achieved. In the meanwhile,
a lot of ``exotic'' features have been discovered to compose
the extremely rich phase diagram of these compounds.
Most recently, theoretical and experimental research has addressed
the issues of the pseudogap onset and of a possible stripe ordering.
Many models have been proposed in order to account at least for some
of the several anomalies in the high temperature superconductivity compounds
(HTSC). Among them, the role of the electron-phonon
({\em el-ph}) coupling has received
alternate fortune. In this contribution we review \hfill the
\begin{table}[h]
\tcaption{Characteristic features of conventional superconductors
compared with the high-$T_c$ materials.}
\end{table}
\begin{center}
\begin{tabular}{l|l}
\hline
\hline
conventional materials  & \hspace{13mm} HTSC\\
\hline
\mbox{}  & \mbox{}  \\
 $T_c^{{\rm max}} \sim 20$  K &\hspace{5mm} $T_c \sim 40 \div 100$ K\\
\mbox{}  & \mbox{}  \\
$\alpha_{T_c} = 0.5$ &\hspace{5mm} $\alpha_{T_c} \sim 0.1 \div 0.8$\\
\mbox{}  & \mbox{}  \\
$\alpha_{m^*} = 0$ &\hspace{5mm} $\alpha_{m^*} \sim -0.6 \div -0.8$\\
\mbox{}  & \mbox{}  \\
$\rho(T) \propto T^5$ &\hspace{5mm} $\rho(T) \propto T$\\
\mbox{}  & \mbox{}  \\
$s$ wave&\hspace{5mm} $d$ wave\\
\mbox{}  & \mbox{}  \\
large bands &\hspace{5mm} narrow bands\\
(high density of &\hspace{5mm} (low density of\\
charge carriers) &\hspace{5mm} charge carriers)\\
\mbox{}  & \mbox{}  \\
phononic pairing  &\hspace{5mm} pairing (?)\\
\mbox{}  & \mbox{}  \\
\hline\hline
\end{tabular}
\vspace{5mm}
\end{center}
main experimental
evidences of a relevant role of the {\em el-ph} interaction on
the superconductive pairing. We discuss these evidences in the context
of the nonadiabatic theory of superconductivity and of the 
normal state.\cite{psg,gps,gpsprl}
The anomalies of the {\em el-ph} phenomenology, which were interpreted
as hints of negligible conventional {\em el-ph} interaction,
acquire now a natural explanation in the nonadiabatic regime.
We also point briefly out analogies and differences between the
cuprate family and the fullerides.

From the point of view of the conventional theory of superconductivity,
described by Migdal-Eliashberg (ME) equations, the high value
of the critical temperature $T_c$ in the HTSC is by itself 
a puzzle. The well-known McMillan formula relates $T_c$ essentially
to two microscopic parameters, $\Omega_b$ and $\lambda_b$, which represent
respectively the energy scale of the intermediate boson and its
coupling strength with electrons. In order to achieve in conventional theory
$T_c$'s as high as $100$ K one should assume an anomalous
large $\lambda_b$, physically prevented by structural distortions,
or alternatively high energy bosons such as electronic excitations.
This latter idea was initially supported by the discovery of a negligible
isotope effect on $T_c$, $\alpha_{T_c}$, at optimal doping.
However, later works found a drastic increase of $\alpha_{T_c}$,
up to $\alpha_{T_c} \simeq 0.8$, as the materials were underdoped.\cite{isot}
Moreover, recent studies reported a finite, large and negative isotope
effect on the electronic mass $\alpha_{m^*} \sim -0.6 \div -0.8$, whereas
$\alpha_{m^*}=0$ is expected in conventional ME theory.\cite{zhao}
These experimental results point out the relevance of {\em el-ph}
scattering in determining superconducting and normal state properties.
The small value of $\alpha_{T_c}$ as well as
the linear dependence on temperature of the resistivity, both peculiarities
of the optimal doping, could be understood better as signature
of some anomalous feature of optimal doping on the top of a phononic pairing
scenario rather than characteristic of the whole phase diagram.
Is $d$-wave symmetry of the superconductive gap compatible with
this outlined picture? Conventional {\em el-ph} pairing does not usually
show any momentum structure, yielding an isotropic $s$-wave superconductivity.
However, it has been shown by different techniques that strong 
electronic correlation induces a momentum structure with a predominance
of forward scattering (small ${\bf q}$'s).\cite{zeyher}
In this situation $d$-wave
symmetry can be favoured with respect to $s$-wave even within a phonon
driven superconductivity.

As it is clear from the above brief overview of experimental results,
{\em el-ph} coupling seems to play a major role in the normal
and superconductive phenomenology, but at the same time it can not be
properly understood within the framework of conventional ME theory.
The new perspective we propose is the nonadiabatic theory of
Fermi liquid.\cite{psg,gps,gpsprl}

Electronic structure in high-$T_c$ materials is characterized
by narrow bands crossing the Fermi level. This situation has two fundamental
consequences with respect to both phononic and electronic scattering:
($i$) on one hand, electronic bands in these compounds can be so narrow
to be of the same order of the phonon energy scale;
($ii$) on the other hand, strong electronic correlation effects
are also definitively important in narrow band systems.
Theoretical works has focused mainly on the second point. In our
opinion, the first one can be equally and even more important.

The adiabatic parameter,
defined as ratio between phonon frequencies and Fermi energy
$\Omega_{ph}/E_F$, can be as large as $\Omega_{ph}/E_F \sim 0.3$ in cuprates.
In this regime Migdal's theorem, on which conventional {\em el-ph}
ME theory rests, breaks down.\cite{migdal} In order to investigate the
nonadiabatic regime, in the past years we have developed
a new theory using a different perturbation approach based on
$\lambda \,\Omega_{ph}/E_F$ instead of $\Omega_{ph}/E_F$.\cite{psg,gps,gpsprl}
A sketch of the Cooper channel interaction in the nonadiabatic
regime is depicted in Fig. 1. 
\begin{figure}[t]
\centerline{\psfig{figure=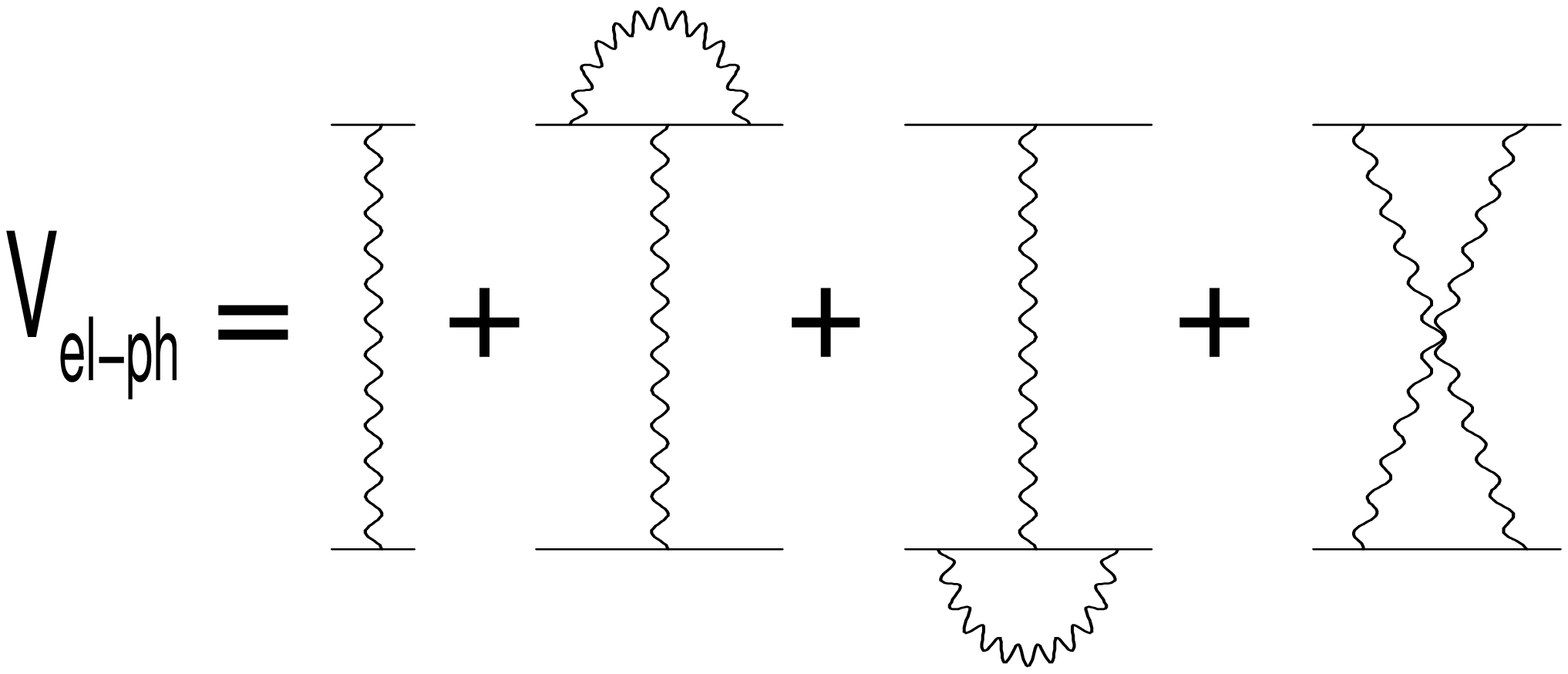,width=9cm,height=3cm,angle=-90}}
\fcaption{Particle-particle interaction in nonadiabatic regime including
corrections beyond Migdal's theorem.}
\end{figure}
A point which we would like to stress
is that nonadiabaticity defines a {\em qualitatively} new theory wherein
{\em el-ph} interaction plays a different role than in
conventional ME theory. In this context, strong correlation acts
as a positive factor with respect to {\em el-ph} coupling.
As we have shown, the forward scattering predominance induced
by the strong electronic correlation selects the phase space where
nonadiabatic effects enhance the effective {\em el-ph} interaction.

In a schematized picture, we can say nonadiabaticity provides a enlarged
scenario where corrections to the conventional ME theory,
arising from the breakdown of Migdal's theorem, can favour or
depress superconducting pairing depending on the degree
of electronic correlation in the system. This latter can be parametrized
by the quantity {\bf $q$}$_c$ representing the momentum selection
induced by the strong correlation.\cite{psg}
More correlated the system smaller {\bf $q$}$_c$.\cite{zeyher}

The issue of a phonon or non-phonon driven superconductivity
is now re-analyz\-ed in the light of the nonadiabatic theory
of superconductivity. The evidences previously discussed
of an ``anomalous'' {\em el-ph} pairing can now be understood
in a natural way in the context of phonon mediated nonadiabatic
superconductivity
sustained by strong electronic correlation.

\begin{itemlist}
\item \underline{$T_c$}: critical temperatures
as large as the experimental ones (up to 100 K) are accompanied in
HTSC by a strong coupling phenomenology (large ratio $2 \Delta/T_c$,
anomalous dip in tunnelling, etc \ldots).
Conventional phonon based ME theory requires an unphysically
large {\em el-ph} coupling ($\lambda \sim 4$) to account for these
features.\cite{combescot} 
In several works we have shown as nonadiabatic corrections
modify the {\em el-ph} interaction and the structure itself
of the Eliashberg equations generalized in nonadiabatic regime.
For small {\bf $q$}$_c$'s (strong correlation) nonadiabatic corrections
enhance the effective {\em el-ph} pairing reproducing high
critical temperatures and strong coupling phenomenology with
reasonable values of $\lambda$ ($\lambda \sim 1$).\cite{benedetti,cgpsu}

\item \underline{$\alpha_{T_c}$}: By simple scaling analyses, the isotope
effect $\alpha_{T_c}$ in ME theory (with no Coulomb repulsion: $\mu=0$)
is shown to be $\alpha_{T_c}=0.5$ for {\em any} $\lambda$.
However, as above discussed, nonadiabatic theory
does not yield just an ``effective'' enhanced {\em el-ph} coupling
$\lambda$, but defines a qualitatively new theory
where a strong coupling phenomenology arises from normal value of $\lambda$.
In similar way,
evaluation of the isotope effect is also deeply different in
nonadiabatic superconductivity. In particular, $\alpha_{T_c}$
shows strong fluctuations ($\alpha_{T_c}\sim 0.2-0.8$ for $\mu=0$) for small
{\bf $q$}$_c$'s, precisely where the enhancement of $T_c$ is the largest,
in agreement with measurement data.\cite{gps,gpsprl}
An additional role can be played by the Van Hove singularity experimentally
observed.\cite{cp}

\item \underline{$\alpha_{m^*}$}: the different structure of the nonadiabatic
equations of superconductivity is reflected also in a finite
and negative isotope effect on the effective electronic mass $m^*$,\cite{gcp}
as experimentally observed.\cite{zhao}
This can be therefore considered as a trademark of nonadiabaticity.
Other possible explanations, as polaron band narrowing or closeness
of a Van Hove singularity, do not seem satisfactory, although
Van Hove singularity is certainly present.

\item \underline{$\rho(T) \propto T$}: one of the most striking features of 
HTSC is the linear dependence of $\rho(T)$
in a wide range of temperature
(up to 1100 K in La214).\cite{allen} At so high temperature 
resistivity is expected to be dominated by phonon scattering.
A linear resistivity with no change of slope suggests
therefore a common, phonon based, scattering
mechanism for high and low temperature. 
As we have shown in Ref. 11,
the linear behaviour can be related to
the presence of a Van Hove singularity located at optimal doping
in the context of nonadiabatic phonon scattering.\cite{cp}

\item \underline{$d$-wave}: it is often argued in literature that $d$-wave
superconductivity is not compatible with phonon pairing
while it points out a spin fluctuation interaction. Reason for
this belief is the structureless {\em el-ph} interaction
of conventional
ME materials which leads to $s$-wave symmetry. HTSC compounds, however,
are characterized by strong correlation inducing an 
important momentum selection with predominance of forward scattering.
In such a situation $d$-wave superconductivity is favoured with respect
to $s$-wave. We find a crossover from $s$- to $d$-wave symmetry
by lowering the momentum selection {\bf $q$}$_c$, or, in other words,
by increasing the rate of electronic correlation.
Nonadiabatic effects are shown to increase both the kind 
of ordering.\cite{pgp}

\item \underline{low charge carrier density}: 
Another puzzling feature of HTSC materials,
as shown by Uemura's plot,\cite{uemura} is the extreme
low density of charge carriers, at least one order of magnitude
lower than in low temperature superconductors.
It is therefore quite surprising that the best superconductors are those with
the poorest number of carriers whereas 
in conventional ME theory $T_c$ increases
with the number of carriers through the density of states.
This inconsistency is solved in natural way
in the nonadiabatic theory of superconductivity.
Small number of carriers leads to small Fermi energies
driving the system into nonadiabatic regime where correction
beyond Migdal's theorem become important.
Moreover, low density of charges implies poor screening of
long-range {\em el-ph} interactions 
(small {\bf $q$}$_c$'s) and predominance of small exchanged momenta.
\end{itemlist}

From the above analysis of different ``exotic'' features of
HTSC materials the fundamental role of {\em el-ph} interaction
appears evident. However, these fature can not be properly explained
in the conventional framework of ME theory. The nonadiabatic
theory of superconductivity and normal state provides a coherent
picture where the above discussed anomalies arise as natural hallmarks
of nonadiabatic effects. The primary actors are two:
on one hand small Fermi energies determine the nonadiabaticity rate
of the system and turn on nonadiabatic corrections due
to the breakdown of Migdal's theorem.
Within this enriched phonon based scenario an additional but
fundamental role is played by the strong electronic correlation
that, through the induced momentum selection, amplifies
the nonadiabatic effects and yields an effective enhanced {\em el-ph}
coupling. This modified {\em el-ph} interaction, generalized in
nonadiabatic regime, defines a new theory
qualitatively different both from the conventional ME one
and from the polaronic picture.
Secondary effects can be also related to supporting actors,
as the Van Hove singularity, that need in any case
to be considered to account for peculiar details on the phase diagram.

All through this contribution, we have mainly focused on the physics
of cuprates, the most studied HTSC compounds by theoretical
and experimental means. However, a look at Uemura's plot, completed
by later materials like the fullerides, suggests a common origin
of superconductivity for the different families of ``HTSC'' materials,
as cuprates, bismuthates, fullerides and heavy fermion systems.\cite{uemura}
All these compounds are characterized by small density of charge carriers
in contrast to conventional ME materials, and nonadiabaticity
stands out as the natural candidate to explain superconductivity in these
systems. The relevant parameter will be the ratio between
intermediate boson frequencies (spin fluctuations in heavy fermions
or phonons in fullerides and bismuthates) and the Fermi energies.
In particular, alkali-doped C$_{60}$ compounds appear, for their
relatively simple phase diagram, as the best materials where 
to check the nonadiabatic Fermi liquid picture.
A detailed study shows how the conventional ME theory can not
explain the experimental data available for these materials
(high values of $T_c$, of $2 \Delta /T_c$, small value of $\alpha_{T_c}$)
while the nonadiabatic theory is able to reproduce the experimental
scenario with quite realistic values of the microscopical
parameters.\cite{cgpsu}
In addition, our proposed description 
is liable to be tested in experimental different ways.
For instance, we predict a finite isotope effect on the 
electronic specific heat\cite{gcp} and on the spin susceptibility\cite{gp}
in fullerides and
more generally in any nonadiabatic superconductor.
An anomalous reduction of $T_c$ by paramagnetic impurity scattering
is also expected.\cite{sgp}
Experimental accuracy for these kind of measurements in nowadays
already available and any 
experimental research along this line is welcome.

\nonumsection{Acknowledgements}
\noindent
E. C. acknowledges the support of the INFM PRA-HTSC project.


\begin{thebibliography}{99}

\bibitem{psg} L. Pietronero, S. Str\"assler and C. Grimaldi,
Phys. Rev. B {\bf 52}, 10516 (1995).

\bibitem{gps}
C. Grimaldi, L. Pietronero and S. Str\"assler,
Phys. Rev. B {\bf 52}, 10530 (1995).

\bibitem{gpsprl}
C. Grimaldi, L. Pietronero and S. Str\"assler,
Phys. Rev. Lett. {\bf 75}, 1158 (1995).

\bibitem{isot}
M.K. Crawford {\em et al.}, Science {\bf 250}, 1390 (1990);

J.P. Franck {\em et al.}, Physica C {\bf 185-189}, 1379 (1991).

\bibitem{zhao}
G. M. Zhao, M. B. Hunt, H. Keller and K. A. M\"{u}ller,
Nature {\bf 385}, 236 (1997).

\bibitem{zeyher}
R. Zeyher and M. Kuli\'c, Phys. Rev. B {\bf 53} 2850 (1996);

M. Grilli and C. Castellani,Phys. Rev. B {\bf 50}, 16880 (1994).

\bibitem{migdal}
A. B. Migdal, Sov. Phys. JETP {\bf 7}, 996 (1958).

\bibitem{combescot}
R. Combescot and G. Varelogiannis,
Europhys. Lett. {\bf 17}, 635 (1992).

\bibitem{benedetti}
P. Benedetti, C. Grimaldi, L. Pietronero and G. Varelogiannis,
Europhys. Lett. {\bf 28}, 351 (1994).

\bibitem{cgpsu}
E. Cappelluti, C. Grimaldi, L. Pietronero, S. Str\"assler and G.A. Ummarino,
condmat/0002221 (2000).

\bibitem{cp}
E. Cappelluti and L. Pietronero,
Europhys. Lett. {\bf 36}, 619 (1996).

\bibitem{gcp}
C. Grimaldi, E. Cappelluti and L. Pietronero,
Europhys. Lett. {\bf 42}, 667 (1998).

\bibitem{allen}
For a review see P.B. Allen, Z. Fisk and A. Migliori,
in {\em Physical Properties of High Temperature Superconductors},
ed. D.M. Ginsberg (World Scientific, Singapore, 1989), p. 213.

\bibitem{pgp}
P. Paci, C. Grimaldi and L. Pietronero,
submitted to Eur. Phys. J. B (2000).

\bibitem{uemura}
Y.J. Uemura {\em et al.}, Phys. Rev. Lett. {\bf 66}, 2665 (1991).

\bibitem{gp}
C. Grimaldi and L. Pietronero,
Europhys. Lett. {\bf 47}, 681 (1999).

\bibitem{sgp}
M. Scattoni, C. Grimaldi and L. Pietronero,
Europhys. Lett. {\bf 47}, 588 (1999).


\end{thebibliography}
\end{document}